\documentclass[prb,preprint,ajp]{revtex4-1} 

\usepackage{amsmath} 

\usepackage{graphicx} 

\usepackage{mathrsfs}

\preprint{\tiny The following article has been accepted by 
the American Journal of Physics. After it is published, it will be 
found at http://scitation.aip.org/ajp.}

\begin{document}

\preprint{\tiny The following article has been accepted by 
the American Journal of Physics. After it is published, it will be 
found at http://scitation.aip.org/ajp.}

\title{Modeling a falling slinky}

\author{R.~C.~Cross}
	\email{cross@physics.usyd.edu.au}
    \affiliation{School of Physics, University of Sydney, NSW 2006, Australia}
\author{M.~S.~Wheatland}
	\email{m.wheatland@physics.usyd.edu.au}
    \affiliation{School of Physics, University of Sydney, NSW 2006, Australia}

\begin{abstract}
A slinky is an example of a tension spring: in an unstretched
state a slinky is collapsed, with turns touching, and a finite
tension is required to separate the turns from this state. 
If a slinky is suspended 
from its top and stretched under gravity and then released, the bottom 
of the slinky does not begin to fall until the top section of the 
slinky, which collapses turn by turn from the top, collides with the 
bottom. The total collapse time $t_c$ (typically $\sim 0.3$\,s 
for real slinkies) corresponds to the time required for a wave front 
to propagate down the slinky to communicate the release of the top 
end. We present a modification to an existing model for a falling 
tension spring~\cite{calkin93} and apply it to data from filmed drops
 of two real slinkies. The modification of the model is the inclusion 
of a finite time for collapse of the turns of the slinky behind the 
collapse front propagating down the slinky during the fall. The new 
finite-collapse time model achieves a good qualitative fit to the 
observed positions of the top of the real slinkies during the measured 
drops. The spring constant $k$ for each slinky is taken to be a free 
parameter in the model. The best-fit model values for $k$ for each 
slinky are approximately consistent with values obtained from measured 
periods of oscillation of the slinkies.
\end{abstract}

\maketitle

\section{Introduction \label{sec:intro}}

The physics of slinkies has attracted attention since their invention
in 1943. Topics of studies include the hanging configuration of the
slinky,~\cite{mak87,sawicki2002} the ability of a slinky to walk 
down stairs,~\cite{hu2010} the modes of oscillation of a vertically 
suspended slinky,~\cite{bowen82,young93} the dispersion of waves 
propagating along slinkies,\cite{blake79,crawford87,vandegrift89}
and the behavior of a vertically stretched slinky when it is 
dropped.~\cite{calkin93,aguirregabiria2007,unruh2011} 

\begin{figure}[h!]
\begin{center}
\scalebox{0.75}{\includegraphics{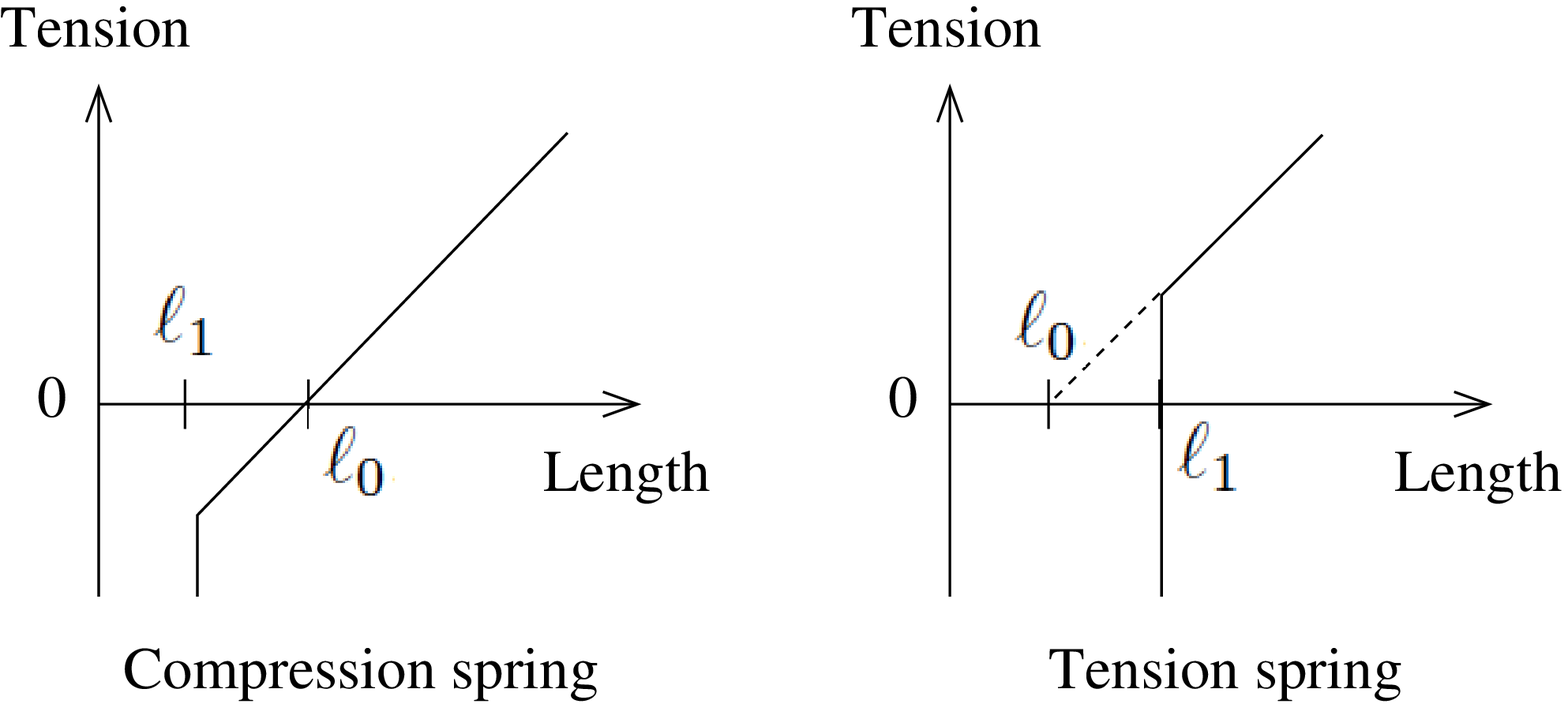}}
\caption{\label{fig:spring_tension}Tension versus length diagrams 
for a compression spring (left) and a tension spring (right). The 
tension in each spring is zero for spring length $\ell_0$ assuming 
Hooke's law applies (this length is not achieved for the tension 
spring). The turns of the spring touch for length $\ell_1$.}
\end{center}
\end{figure}

Slinkies are examples of tension springs, i.e.\ springs which may
be under tension according to Hooke's law, but not compression. 
Unstretched slinkies have a length $\ell_1$ at which the turns are in 
contact, and a finite tension $f_1$ is needed to separate the turns 
from this state. They collapse to this state if not stretched by an 
external force. This may be contrasted with a ``compression spring,'' 
which can be under tension or compression according to Hooke's law. 
Compression 
springs have an unstretched length $\ell_0$ at which the turns are not
in contact, and the tension is zero. They may be compressed to a 
length $\ell_1$ at which the turns are in contact, and obey Hooke's
law during this compression. Fig.~\ref{fig:spring_tension} shows
tension versus length diagrams for uniform extensions of the two types 
of springs.

The vertically falling slinky, mentioned above, exhibits interesting 
dynamics which depend on the slinky being a tension 
spring.~\cite{calkin93} A falling compression spring exhibits periodic 
compressions and rarefactions, as longitudinal waves propagate 
along the spring length. A falling tension spring collapses to 
the length $\ell_1$ during a fall, assuming the spring is released in 
an initially stretched state (with length $>\ell_1$). 

If a slinky is hanging vertically under gravity from its top (at rest) 
and then released, the bottom of the slinky does not start to move 
downwards until the collapsing top section collides with the bottom.  
Figure~\ref{fig:slinky_video} illustrates this peculiar effect for a 
plastic rainbow-colored slinky;
this figure shows a succession of frames extracted from a high-speed 
video of the fall of the slinky.~\cite{wheatland_cross12} 
The continued suspension of the bottom of the slinky after 
release is somewhat counter-intuitive and very intriguing---a 
recent YouTube video showing the effect with a falling slinky has 
received more than 800,000 views.~\cite{muller_cross10} The physical explanation is 
straightforward: the collapse of tension in the slinky occurs from the
top down, and a finite time is required for a wave front to propagate 
down the slinky communicating the release of the top. 

The basic wave physics behind this behavior follows from the equation 
of motion for a falling (or suspended) compression spring\cite{edwards72} 
\begin{equation}\label{eqn:eom_dim}
m\frac{\partial^2 x}{\partial t^2}=k\frac{\partial^2x}{\partial \xi^2}
+mg,
\end{equation}
which applies to a tension spring when the turns are separated.
In this equation $x(\xi,t)$ is the vertical location of a point 
along the spring at time $t$, $m$ is the total spring mass, and $k$
is the spring constant. The (dimensionless) coordinate $\xi$ describes the mass
distribution along the spring, such that $dm=m\,d\xi$ is the
increment in mass associated with an increment in $\xi$, with 
$0\leq \xi\leq 1$. Thus, for a spring with $N$ turns, the end of turn $i$ 
corresponds to $\xi_i=i/N$ and is located at position 
$x_i=x(\xi_i,t)=x(i/N,t)$ at time $t$. Equation~(\ref{eqn:eom_dim}) is an 
inhomogenous wave
equation. If the spring is falling under gravity, then in a coordinate 
system falling with the center of mass of the spring 
($x^{\prime}=x-\frac{1}{2}gt^2$), the equation of motion is the usual 
wave equation
\begin{equation}\label{eqn:eom_com_dim}
m\frac{\partial^2 x^{\prime}}{\partial t^2}=
k\frac{\partial^2x^{\prime}}{\partial \xi^2}.
\end{equation}
Equation~(\ref{eqn:eom_com_dim}) implies that waves in the mass distribution
(turn spacing) propagate along the length of the spring in a 
characteristic time $t_p=\sqrt{m/k}$. This accounts for the
periodic rarefactions and compressions of a compression spring during
a fall, and for the propagation of the wave front ahead of the 
collapsing turns in a falling tension spring.

In this paper we present a new detailed model for the fall of a 
slinky, which improves on past models
by taking into account the finite time for collapse of the turns of
the slinky behind the wave front. In Sec.~\ref{sec:collapse}
we explain the need for this refinement in the modeling, and we
present the details of the new model in Sec.~\ref{sec:model}. 
The new model is compared with the behavior of two real falling 
slinkies in Sec.~\ref{sec:appl}, and we discuss our conclusions
in Sec.~\ref{sec:conc}.

\section{The collapse of the turns at the top of the 
slinky\label{sec:collapse}}

A detailed description of the dynamical behavior of a falling tension
spring requires solution of the equations of motion for mass elements
along the slinky that are subject to gravity and local spring 
forces, taking into account the departure from Hooke's law encountered 
when slinky turns come into contact. Because of the complexity of this
modeling, past efforts involve specific
approximations.~\cite{calkin93,aguirregabiria2007,unruh2011}

For a mass element $m\Delta \xi$ at a location $\xi_i$ on the slinky, 
the equation of motion is\cite{edwards72} 
\begin{equation}\label{eqn:eom_gen_dim}
\begin{split}
m\Delta \xi\left.\frac{\partial^2 x}{\partial t^2}\right|_{\xi_i,t}
  &=f(\xi_i+\Delta \xi,t)-f(\xi_i,t)+m\Delta \xi g \\
  &=\Delta \xi\left.\frac{\partial f}{\partial \xi}\right|_{\xi_i,t}
  +m\Delta\xi g,
\end{split}
\end{equation}
where $f(\xi,t)$ is the tension force 
at $\xi$. Equation~(\ref{eqn:eom_gen_dim}) applies to all points
except the top and bottom of the slinky, where the tension is
one sided.
When slinky turns are separated at a point along the 
slinky, the tension is given by Hooke's law in the 
form\cite{calkin93}
\begin{equation}\label{eqn:hookes_tension}
f(\xi,t)=k\left(\frac{\partial x}{\partial \xi}-\ell_0\right),
\end{equation}
where $\partial x/\partial \xi$ describes the local extension of the 
slinky, and $\ell_0$ corresponds to a slinky length at which the tension
would be zero, assuming a Hooke's law relation for all values of the 
local extension. Substituting Eq.~(\ref{eqn:hookes_tension}) into
Eq.~(\ref{eqn:eom_gen_dim}) leads to the inhomogenous wave
equation~(\ref{eqn:eom_dim}).

For a tension spring the length $\ell_0$ cannot be reached because
there is a minimum length
\begin{equation}\label{eqn:collapsed}
\frac{\partial x}{\partial \xi}=\ell_1>\ell_0
\end{equation}
that corresponds to the spring coils being in contact with each other.
At this point, the minimum tension
\begin{equation}\label{eqn:min_tension}
f_1=k(\ell_1-\ell_0) 
\end{equation}
is achieved and the tension is
replaced by a large (infinite) compression force as the collapsed turns resist
further contraction of the slinky (see Fig.~\ref{fig:spring_tension}).
This non-Hooke's law behavior is met when turns collapse at
the top of the falling slinky and the description of this process
complicates the modeling. 

A simpler, approximate description of the dynamical collapse of the
top of the slinky is to assume a functional form for the position-mass
distribution $x(\xi,t)$ during the collapse, and then impose
conservation of momentum to ensure physical time evolution.
Calkin~\cite{calkin93} introduced this semi-analytic approach and
specifically assumed a distribution corresponding to slinky turns
collapsing instantly behind a downward propagating wave front located 
a mass fraction $\xi_c=\xi_c(t)$ along the slinky at time $t$ after the
release. The turns  of the slinky at the front instantly assume a
configuration with a minimum tension, so that
Eq.~(\ref{eqn:collapsed}) applies for all points behind the front at a
given time
\begin{equation}\label{eqn:free_bc_top}
f(\xi < \xi_c,t)=f_1. 
\end{equation} 
For points ahead of the front ($\xi > \xi_c$) the tension is the same as in the hanging slinky.
The location of the front at time $t$ is obtained by requiring 
that the total momentum of the collapsing slinky matches the impulse
imparted by gravity up to that time. (The modeling is presented
in detail in Sec.~\ref{sec:model_basic}.)
The Calkin model has also been derived in
solving the inhomogenous wave equation~(\ref{eqn:eom_dim}) subject
to the boundary condition given by
Eq.~(\ref{eqn:free_bc_top}).~\cite{aguirregabiria2007,unruh2011}

With real falling slinkies the collapse of turns behind the front
takes a finite time. Figure~\ref{fig:collapse_diagram} illustrates the
process of collapse of a real slinky using data extracted from a
high-speed video of a fall (this data is discussed in more detail in
Sec.~\ref{sec:appl_data}.) The upper panel of
Fig.~\ref{fig:collapse_diagram} shows the position of the top (blue
circles), of turn eight (black $+$ symbols), and of turn ten (red
$\times$ symbols) versus time, for the first $0.2$\,s of the fall. The
vertical position is shown as negative in the downward direction
measured from the initial position of the top [which corresponds to
$-x$ in terms of the notation of Eq.~(\ref{eqn:eom_com_dim})]. The
upper panel shows that turns 8 and 10 of the slinky remain at rest
until the top has fallen some distance, and then turn eight begins to
fall before turn ten. The lower panel shows the spacing of turns eight
and ten versus time. The two turns change from the initial stretched
configuration to the final collapsed configuration in  
$\sim 0.1$\,s.

This paper presents a method of solution of Eq.~(\ref{eqn:eom_dim})
which adopts the approximate approach of Calkin, but includes a finite time for 
collapse of the turns. We assume a linear profile for the 
decay in tension behind the wave front propagating down the slinky,
which provides a more realistic description of the slinky collapse.

\section{Modeling the fall of a slinky\label{sec:model}}

In Secs.~\ref{sec:model_hanging} and~\ref{sec:model_basic} we 
reiterate the Calkin~\cite{calkin93} model for a hanging slinky as 
a tension spring, and for the fall of the tension spring. In 
Sec.~\ref{sec:model_mod} the new model for the fall of the slinky is 
presented.

\subsection{The hanging slinky \label{sec:model_hanging}}

For hanging slinkies it is generally observed that the top section of 
the slinky has stretched turns, and a small part at the bottom has
collapsed turns.~\cite{mak87} Assuming mass 
fractions $\xi_1$ and $1-\xi_1$ of the slinky with stretched and 
collapsed turns, respectively, the number of collapsed turns $N_c$ 
is related to the total number of turns $N$ by
\begin{equation}\label{eqn:nc}
N_c=(1-\xi_1)N.
\end{equation}
A hanging slinky such that the turns just touch
at the bottom would have $\xi_1=1$.

The position $X=X (\xi )$ of points along the stretched part of the
stationary hanging slinky is described by setting 
$\partial^2 x/\partial t^2=0$ in Eq.~(\ref{eqn:eom_dim}) 
and integrating from $\xi=0$ to $\xi=\xi_1$ with the boundary 
conditions
\begin{equation}
X(\xi=0)=0\quad \mbox{and} \quad
\left.\frac{\partial X}{\partial \xi}\right|_{\xi=\xi_1}=\ell_1,
\end{equation}
corresponding to the fixed location of the top of the slinky, and
the spacing of collapsed turns at the bottom of the slinky, respectively.
The position of points in the collapsed section at the bottom is 
obtained by integrating Eq.~(\ref{eqn:collapsed}) from $\xi=\xi_1$
to $\xi=1$, with the boundary value $X(\xi_1)$ matching the result
obtained by the first integration.  Carrying out these calculations gives
\begin{equation}
X(\xi )=
\begin{cases}
\ell_1\xi +\displaystyle{\frac{mg}{k}}\left(\xi_1-\frac{1}{2}\xi\right)\xi, & \mbox{for }0\leq \xi\leq \xi_1\\
\ell_1\xi+\displaystyle{\frac{mg}{2k}}\xi_1^2, & \mbox{for }\xi_1\leq\xi\leq 1.
\end{cases}
\label{eqn:x0}
\end{equation} 
The total length of the slinky in this configuration is
\begin{equation}\label{eqn:xb0}
X_{\mathrm{B}}=X(1)=\ell_1+\frac{mg}{2k}\xi_1^2,
\end{equation}
where B refers to the bottom of the slinky,
and the center of mass is at
\begin{equation}\label{eqn:xcom0}
X_{\mathrm{com}}=\int_0^1 X(\xi )\,d\xi = \frac{1}{2}\ell_1
  +\frac{mg}{2k}
  \left(1-\frac{1}{3}\xi_1\right)\xi_1^2.
\end{equation}

The left panel of Fig.~\ref{fig:slinky_hang_fall} illustrates the model 
slinky in the hanging configuration. The slinky is drawn as a
helix with a turn spacing matching $X(\xi )$, for model parameter 
values typical of real slinkies (detailed modeling of real slinkies
is presented in Sec.~\ref{sec:appl}). The chosen parameters are: 
$80$ total turns ($N=80$), slinky mass $m=200$\,g, hanging length 
$X_{\mathrm{B}}=1$\,m, collapsed length $\ell_1=60$\,mm, slinky 
radius $30$\,mm, and $10\%$ of the slinky mass collapsed 
at the bottom when hanging ($\xi_1=0.9$). The light gray (green online)
section of the slinky at the bottom is the collapsed section, and the
dot (red online) is the location of the center-of-mass of the slinkey
(given by Eq.~(\ref{eqn:xcom0}) in the left panel).

\subsection{The falling slinky with instant collapse of 
turns\label{sec:model_basic}}

We assume the slinky is released at $t=0$ and the turns collapse 
from the top down behind a propagating wave front. In the model with
instant collapse~\cite{calkin93} the process is completely described 
by the location $\xi_c=\xi_c(t)$ 
of the front at time $t$. The slinky is collapsed where 
$0\leq \xi<\xi_c $ but is still in the initial state where 
$\xi_c<\xi\leq 1$. 
The position of points in the collapsed 
section of the slinky, behind the front, is obtained by integrating 
Eq.~(\ref{eqn:collapsed}) and matching to the boundary condition
\begin{equation}
x (\xi=\xi_c,t)=X(\xi_c),
\end{equation}
to get
\begin{equation}\label{eqn:x_calkin_top}
x (\xi,t)=\ell_1\xi
  +\frac{mg}{k}\xi_c\left(\xi_1-\frac{1}{2}\xi_c\right)
\end{equation}
for $0\leq \xi\leq\xi_c$.
The lower part of the slinky ($\xi_c\leq\xi\leq 1$) has positions
described by Eq.~(\ref{eqn:x0}).

The motion of the slinky after release follows from Newton's second 
law. The collapsed top section has a velocity given by the derivative 
of Eq.~(\ref{eqn:x_calkin_top})
\begin{equation}\label{eqn:v_calkin}
\frac{\partial x(\xi,t)}{\partial t}=\frac{mg}{k}\left(\xi_1-\xi_c\right)
  \frac{d\xi_c}{dt},
\end{equation}
and the mass of this section is $m\xi_c$. The rest of the slinky is
stationary so the total momentum of the slinky is 
obtained by multiplying Eq.~(\ref{eqn:v_calkin}) by the mass 
$m\xi_c$. Setting the momentum equal to the net impulse $mgt$ due to 
gravity on the slinky at time $t$ gives
\begin{equation}\label{eq:eom_basic}
\xi_c\left(\xi_1-\xi_c\right)\frac{d\xi_c}{dt}=\frac{k}{m}t,
\end{equation}
which can be directly integrated to give
\begin{equation}\label{eqn:cubic_calkin}
\xi_c^2\left(\xi_1-\frac{2}{3}\xi_c\right)=\frac{k}{m}t^2.
\end{equation}
At a given time Eq.~(\ref{eqn:cubic_calkin}) is a cubic in $\xi_c$; 
the first positive root to the cubic defines the location of the 
collapse front at that time. The total collapse time $t_c$---the time
for the collapse front to reach the bottom, collapsed section---is 
defined by $\xi_c(t_c)=\xi_1$, and from Eq.~(\ref{eqn:cubic_calkin}) it
follows that
\begin{equation}\label{eqn:tc}
t_c=\sqrt{\frac{m}{3k}\xi_1^3}\,.
\end{equation}

Equation~(\ref{eqn:cubic_calkin}), together with Eqs.~(\ref{eqn:x0}) 
and~(\ref{eqn:x_calkin_top}), defines the location $x(\xi,t)$ of all 
points on the slinky for $t<t_c$. The center of mass falls from rest
with acceleration $g$ and so has location
\begin{equation}\label{eqn:xcom_calkin_collapse}
x_{\mathrm{com}}(t)=X_{\mathrm{com}}+\frac{1}{2}gt^2,
\end{equation}
where $X_{\mathrm{com}}$ is given by Eq.~(\ref{eqn:xcom0}).

Figure~\ref{fig:basic_model} shows solution of the instant-collapse
model with the typical slinky parameters used in
Fig.~\ref{fig:slinky_hang_fall}.~\cite{wheatland_cross12} The upper
panel shows the positions of the top (upper solid curve, blue online),
center-of-mass (middle solid curve, red online), and bottom (lower
solid curve, red online) of the slinky versus time.
Position is negative in the  downward direction
so the upper (blue) curve corresponds to the model expression
$-x(0,t)$.  The position of the front versus time is
indicated by the dashed (black) curve. The lower panel shows the
velocity of the top of the slinky versus time (solid curve, blue online) and in
both panels the total collapse time $t_c$ is indicated by the vertical
dashed (pink) line.   For the typical slinky parameters used, the spring
constant is $k=0.84$\,N/m and the collapse time is $t_c\approx 0.24$\,s.

Figure~\ref{fig:basic_model} illustrates a number of unusual features of
the model. For example, the initial velocity of the top is non-zero---a
consequence of the assumption of instant collapse at the wave front. 
From Eqs.~(\ref{eqn:v_calkin}) and~(\ref{eqn:cubic_calkin}) 
the initial velocity of the top is 
\begin{equation}
v_{T}^0=-\left.\frac{\partial x}{\partial t}\right|_{\xi=0}
  =-g\sqrt{\frac{m\xi_1}{k}}\approx -4.5\,\mathrm{m/s}.
\end{equation}
The acceleration of the top at time $t=0$ must be infinite to produce
a finite initial velocity. The acceleration of the top of the slinky just
after $t=0$ is positive,  i.e.\ in the upwards direction, so the top
of the slinky falls more  slowly with time.  From
Eqs.~(\ref{eqn:v_calkin}) and~(\ref{eqn:cubic_calkin}) the
limiting value of the acceleration as $t\rightarrow 0$ is 
\begin{equation}
a_{T}^0=-\left.\frac{\partial^2 x}{\partial t^2}\right|_{\xi=0}
=\frac{g}{3\xi_1}\approx 3.6\,\mathrm{m/s}^2.
\end{equation}
The acceleration of the top becomes negative (downwards) after the 
collision of the top and bottom sections, when the whole slinky falls 
with acceleration $-g$. At the collapse time when the top section 
impacts the bottom section there is an impulsive collision causing a 
discontinuous jump in the velocity.

\subsection{The falling slinky with a finite time for
collapse of turns\label{sec:model_mod}}

The instant-collapse model requires an unphysical instant change in
the angle of the slinky turns behind the collapse front, as
discussed in Secs.~\ref{sec:collapse} and~\ref{sec:model_basic}.
This affects the positions of all turns of the slinky as a function of
time behind the front. To model the positions of the turns of a real 
collapsing slinky it is necessary to modify the model.

\begin{figure}[h!]
\begin{center}
\scalebox{0.85}{\includegraphics{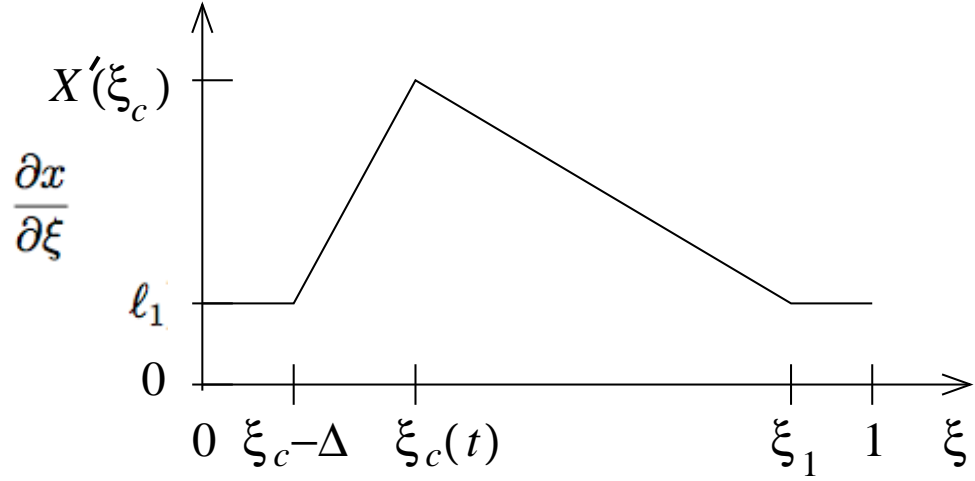}}
\caption{\label{fig:linear_relaxation}The gradient
$\partial x/\partial \xi$, which describes the local slinky extension, 
versus mass density $\xi$ in the finite-collapse-time model.
The tension defined
by this profile declines linearly behind the wave front
[located at $\xi_c (t)$] from a value matching
the tension in the hanging slinky at the front, to the minimum 
tension value $f_1=k(\ell_1-\ell_0)$ at $\xi=\xi_c-\Delta$. 
Ahead of the front the tension is unchanged from that in the 
hanging slinky.}
\end{center}
\end{figure}

The lower panel of Fig.~\ref{fig:collapse_diagram} indicates that 
the spacing between the turns of the slinky decreases approximately
linearly with time during the collapse. Hence we modify the model 
in Sec.~\ref{sec:model_mod} to include
a linear profile for the decay in tension behind the collapse front 
propagating down the slinky, as a function of mass fraction $\xi$.
The tension is assumed to be given by Eq.~(\ref{eqn:hookes_tension})
with
\begin{equation}\label{eqn:collapsing1}
\frac{\partial x}{\partial \xi}=\left[X^{\prime}(\xi_c)-\ell_1\right]
  \left(1+\frac{\xi-\xi_c}{\Delta}\right)+\ell_1
  \quad\mathrm{for}~ \mathrm{max}(0,\xi_c-\Delta)\leq\xi \leq\xi_c,
\end{equation}
where $X=X(\xi )$ is given by Eq.~(\ref{eqn:x0}), and the prime 
denotes differentiation with respect to the parameter $\xi$.  In
this equation, $\Delta$ is a parameter that governs
the distance over which the tension decays back to its minimum
value $f_1$.
Figure~\ref{fig:linear_relaxation} illustrates the local slinky extension 
at time $t$ as described by Eq.~(\ref{eqn:collapsing1}).
Behind the front at $\xi_c(t)$ the extension decreases 
linearly as a function of $\xi$, returning to the minimum value 
$\ell_1$ over the fixed mass fraction $\Delta$. Ahead of the front 
the extension is the same as for the hanging slinky.

Equation~(\ref{eqn:collapsing1}) 
replaces Eq.~(\ref{eqn:collapsed}) for the section of the 
slinky behind the collapse front and provides a simple, approximate 
description of a finite collapse time for the turns behind the front.
The limit $\Delta \rightarrow 0$ in the new model recovers the 
instant-collapse model. 

Using Eq.~(\ref{eqn:x0}) to evaluate the gradient in 
Eq.~(\ref{eqn:collapsing1}) gives
\begin{equation}\label{eqn:collapsing2}
\frac{\partial x}{\partial \xi}=\frac{mg}{k}\left(\xi_1-\xi_c\right)
  \left(1+\frac{\xi-\xi_c}{\Delta}\right)+\ell_1
  \quad\mathrm{for}~ \mathrm{max}(0,\xi_c-\Delta)\leq\xi \leq\xi_c.
\end{equation}
Integrating Eq.~(\ref{eqn:collapsing2}) and imposing the boundary
condition $x(\xi_c)=X(\xi_c)$ using Eq.~(\ref{eqn:x0}) gives
\begin{equation}\label{eqn:collapsed3}
x=\frac{mg}{k}\left[\left(\xi_1-\xi_c\right)
  \left(1-\frac{1}{\Delta}\xi_c+\frac{1}{2\Delta}\xi\right)\right]\xi
  +\ell_1\xi
  +\frac{mg}{2k}\xi_c^2
  \left(1+\frac{\xi_1-\xi_c}{\Delta}\right)
\end{equation}
for $\mathrm{max}(0,\xi_c-\Delta)\leq\xi \leq \xi_c$.
If $\xi_c>\Delta$, there is a completely collapsed
section at the top of the slinky. The mass density in this section
is obtained by integrating Eq.~(\ref{eqn:collapsed}) and matching to 
the value $x(\xi_c-\Delta)$ given by Eq.~(\ref{eqn:collapsed3}),
leading to
\begin{equation}\label{eqn:collapsed4}
x=\ell_1\xi
  +\frac{mg}{k}\left[\xi_c\left(\xi_1-\frac{1}{2}\xi_c\right)
  -\frac{1}{2}\Delta\left(\xi_1-\xi_c\right)\right]
  \quad\mathrm{for}~ 0\leq \xi\leq \mathrm{max}(0,\xi_c-\Delta).
\end{equation}
Equations~(\ref{eqn:collapsed3}) and~(\ref{eqn:collapsed4}) are the 
counterparts to Eq.~\eqref{eqn:x_calkin_top} in the instant-collapse 
model. In the limit $\Delta \rightarrow 0$, Eq.~(\ref{eqn:collapsed4})
is the same as Eq.~(\ref{eqn:x_calkin_top}).

The motion of the slinky in the new model is determined in the 
same way as for the instant-collapse model.
The velocity of the top section of the slinky prior to the 
complete collapse of the top is obtained by differentiating 
Eqs.~(\ref{eqn:collapsed3}) and~(\ref{eqn:collapsed4}) to get
\begin{equation}\label{eqn:v_mod1}
\begin{split}
\frac{\partial x}{\partial t}=
  \frac{mg}{k}&\left[
  \left(1 +\frac{\xi_1-3\xi_c/2}{\Delta}\right)\xi_c
  -\left(1+\frac{\xi_1-2\xi_c+\xi/2}{\Delta}\right)\xi\right]
  \frac{d\xi_c}{dt}, \\
  &\quad\quad\mathrm{for}~ \mathrm{max}(0,\xi_c-\Delta)\leq\xi \leq\xi_c,
\end{split}
\end{equation}
and
\begin{equation}\label{eqn:v_mod2}
\frac{\partial x}{\partial t}=
  \frac{mg}{k}\left(\xi_1-\xi_c+\frac{1}{2}\Delta\right)
  \frac{d\xi_c}{dt},
  \quad\mathrm{for} ~0\leq \xi\leq \mathrm{max}(0,\xi_c-\Delta).
\end{equation}
Equation~(\ref{eqn:v_mod2}) is the counterpart to Eq.~(\ref{eqn:v_calkin}).
The total momentum of the slinky is given by
\begin{equation}
\label{eqn:p_mod_def}
p=m\int_{0}^{\xi_c}\frac{\partial x}{\partial t}d\xi\,,
\end{equation}
and using Eqs.~(\ref{eqn:v_mod1}) and~(\ref{eqn:v_mod2}) to evaluate
the integral gives
\begin{equation}
\label{eqn:p_mod1}
p=\frac{m^2g}{2k}\xi_c^2
  \left(1+\frac{\xi_1-4\xi_c/3}{\Delta}\right)
    \frac{d\xi_c}{dt}
    \quad\mathrm{if} ~\xi_c\leq\Delta,
\end{equation}
and
\begin{equation}
\label{eqn:p_mod2}
p=\frac{m^2g}{k}\left[
  \xi_c\left(\xi_1-\xi_c\right)+\Delta
  \left(\xi_c-\frac{1}{2}\xi_1-\frac{1}{6}\Delta\right)
  \right]
    \frac{d\xi_c}{dt}
\quad\mathrm{if} ~\xi_c\geq \Delta.
\end{equation}
Setting Eqs.~(\ref{eqn:p_mod1}) and~(\ref{eqn:p_mod2}) 
equal to the total impulse $mgt$ on the slinky up to time $t$ gives
equations defining the location $\xi_c(t)$ of the front at
time $t$:
\begin{equation}\label{eqn:eom_mod1}
\frac{1}{2}\xi_c^2\left(1+\frac{\xi_1-\frac{4}{3}\xi_c}{\Delta}\right)
    \frac{d\xi_c}{dt}
  =\frac{kt}{m} \quad\mathrm{if} ~\xi_c\leq\Delta, 
\end{equation}
and
\begin{equation}\label{eqn:eom_mod2}
\left[
  \xi_c\left(\xi_1-\xi_c\right)+\Delta
  \left(\xi_c-\frac{1}{2}\xi_1-\frac{1}{6}\Delta\right)
  \right]
    \frac{d\xi_c}{dt}
  =\frac{kt}{m}\quad\mathrm{if} ~\xi_c\geq\Delta,
\end{equation}
which are the counterparts to Eq.~(\ref{eq:eom_basic}) in the 
instant-collapse model. 
Equations~(\ref{eqn:eom_mod1}) and~(\ref{eqn:eom_mod2}) may be integrated
with respect to $\xi_c$, leading to
\begin{equation}\label{eqn:cubic_mod1}
\xi_c^3\frac{\Delta+\xi_1-\xi_c}{3\Delta}
  =\frac{kt^2}{m}
  \quad\mathrm{if} ~\xi_c\leq\Delta,
\end{equation}
and
\begin{equation}\label{eqn:cubic_mod2}
  \xi_c^2\left(\xi_1-\frac{2}{3}\xi_c\right)
  -\Delta \left(\xi_1-\xi_c\right)\left(\xi_c-\frac{1}{3}\Delta\right)
  =\frac{kt^2}{m}
  \quad\mathrm{if} ~\xi_c\geq\Delta,
\end{equation}
which are the
counterparts to Eq.~(\ref{eqn:cubic_calkin}) that
defines the location of the front in the instant-collapse model.

The total collapse time $t_c$ for the slinky (the time for the front 
to reach $\xi_1$) is obtained by setting $\xi_c=\xi_1$ in 
Eq.~(\ref{eqn:cubic_mod2}).  Interestingly,
the result is unchanged from the instant-collapse case and is given 
by Eq.~(\ref{eqn:tc}). A second time scale relevant for the model
is the time $t_{\mathrm{lin}}$ for the top of the slinky to undergo 
the initial linear collapse (for times $t>t_{\mathrm{lin}}$ there are
completely collapsed turns at the top of the slinky). This is obtained
by setting $\xi_c=\Delta$ in Eq.~(\ref{eqn:cubic_mod1}) to get
\begin{equation}\label{eqn:tlin}
t_{\mathrm{lin}}=\sqrt{\frac{m\xi_1}{3k}}\Delta.
\end{equation}

\begin{figure}[h!]
\begin{center}
\scalebox{0.7}{\includegraphics{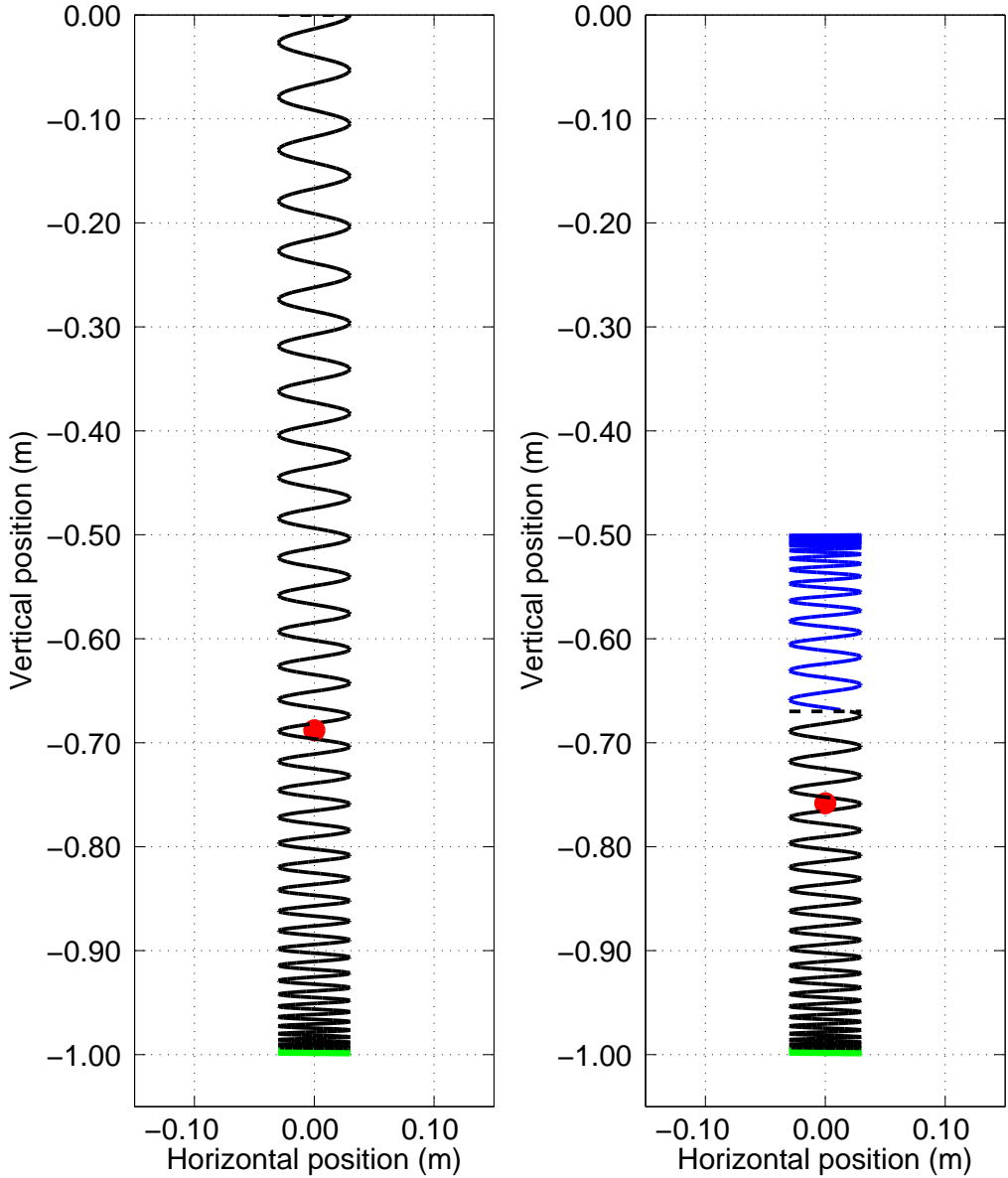}}
\caption{\label{fig:slinky_hang_fall}Left panel: the model for a 
hanging slinky, with the slinky represented as a helix with turn
spacing matching $X(\xi )$. Typical 
slinky parameters are used. The dot in each panel indicates the
location of the center of mass, and the light gray (green online) part of the slinky
at the bottom is the collapsed section.
Right panel: the finite-collapse-time model for the slinky during 
the fall at time $t=t_c/2$. The top dark gray (blue online) section of the 
slinky is the section undergoing collapse, above the 
downward-propagating collapse front indicated by a dashed 
horizontal line.}
\end{center}
\end{figure}

\begin{figure}[h!]
\begin{center}
\scalebox{0.7}{\includegraphics{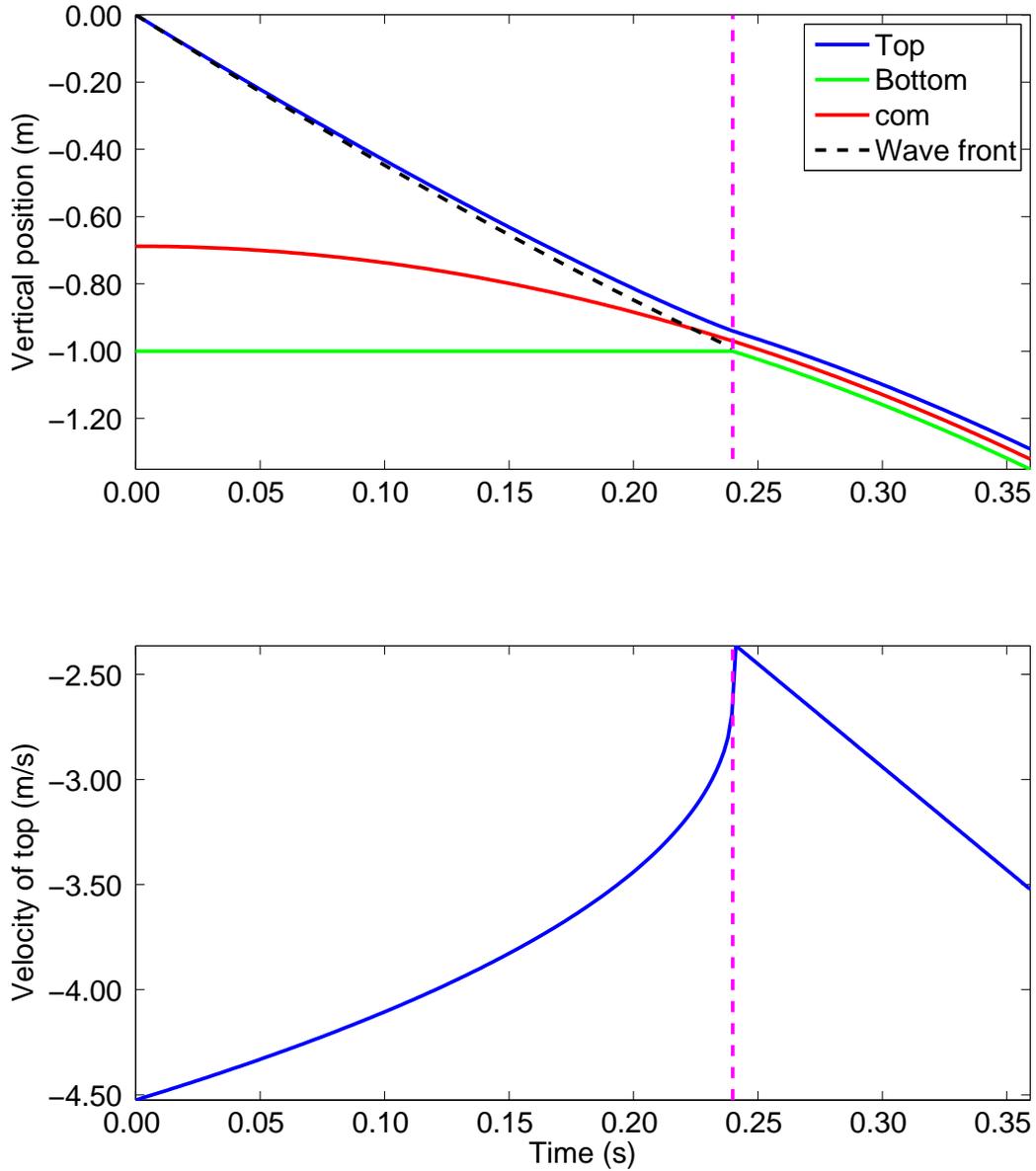}}
\caption{\label{fig:basic_model}The instant-collapse model~\cite{calkin93}
for a falling slinky using parameters typical of a real slinky.
Upper panel: position versus time of the slinky top (upper solid curve),
center-of-mass (middle solid curve), bottom (lower solid curve),
and wave front initiating collapse (dashed curve).  Position is negative downwards 
in this figure. Lower panel: velocity of the slinky top
versus time.  The total collapse time
$t_c$ is shown as the vertical dashed line in both panels.}
\end{center}
\end{figure}

\begin{figure}[h!]
\begin{center}
\scalebox{0.7}{\includegraphics{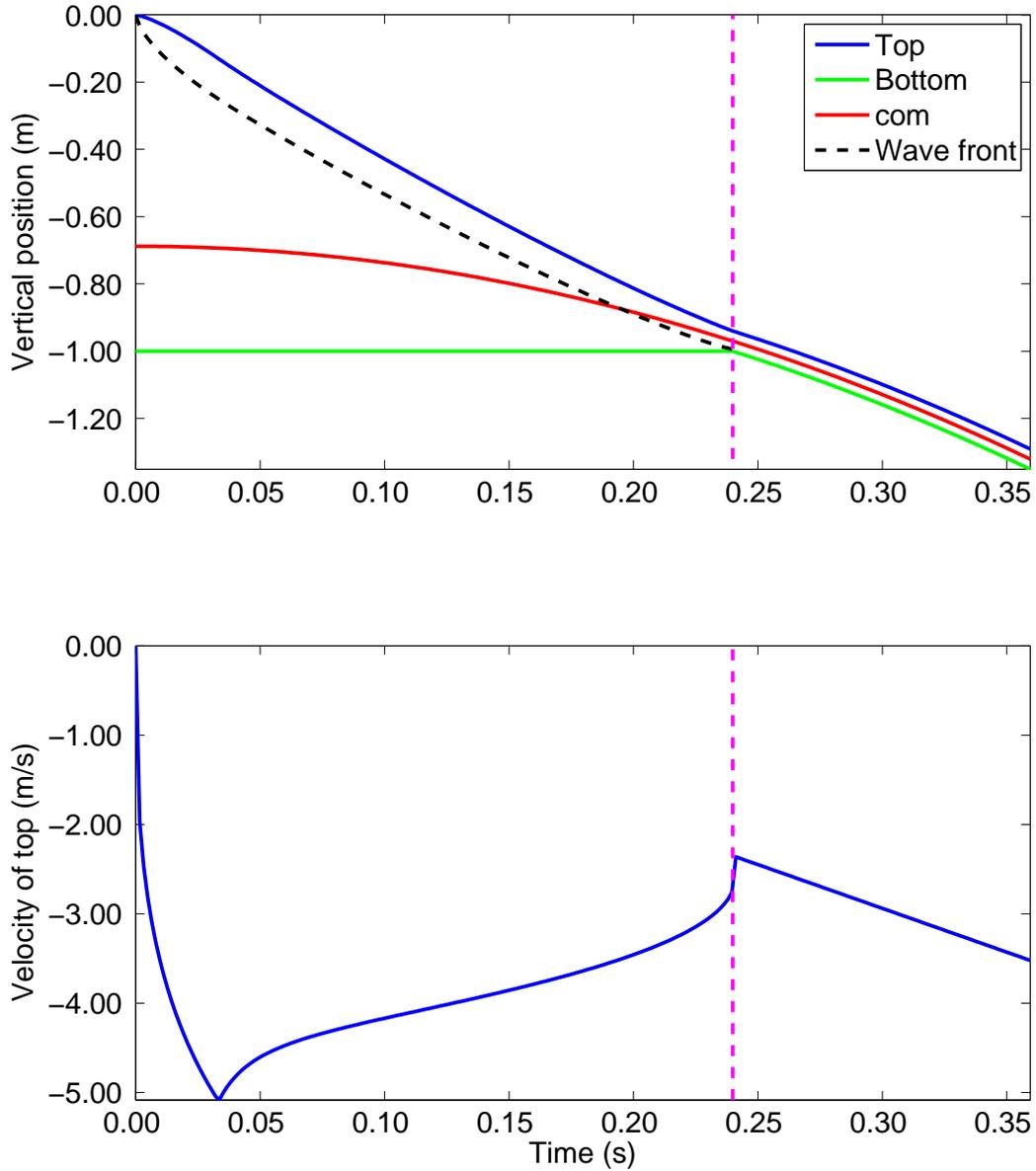}}
\caption{\label{fig:mod_model}The finite-collapse-time model for a falling slinky
using the same slinky parameters 
as in Fig.~\ref{fig:basic_model}. The collapse of the model slinky 
is assumed to occur via a linear decay in tension over ten turns of 
the slinky.
Upper panel: position versus time of 
the slinky top (upper solid curve),
center-of-mass (middle solid curve), bottom (lower solid curve),
and wave front initiating collapse (dashed curve).  Position is negative downwards 
in this figure. Lower panel: velocity of the slinky top
versus time.  The total collapse time
$t_c$ is shown as the vertical dashed line in both panels.}
\end{center}
\end{figure}

Figure~\ref{fig:mod_model} shows solution of the finite-collapse
time model
for the typical parameters used in Figs.~\ref{fig:slinky_hang_fall} 
and~\ref{fig:basic_model}, and with a value of $\Delta$ chosen to match 
10 turns of the 80-turn slinky ($\Delta=10/80=0.125$). The layout of the 
figure is the same as for Fig.~\ref{fig:basic_model}. The position 
versus time of the top of the
slinky (upper solid curve in the upper panel) is very similar to that in the 
instant-collapse model, but the top initially accelerates downwards 
from rest 
rather than having an initial non-zero velocity. The location of the
front versus time (dashed curve in the upper panel) is 
significantly different to that shown in 
Fig.~\ref{fig:basic_model}, and comparison of this curve and the 
position of the top of the slinky shows the effect of the finite 
time for turns to collapse behind the front. A specific feature of
the motion of the front is that the initial velocity of the front is
infinite (the dashed curve has a vertical slope at $t=0$). 
The lower panel of Fig.~\ref{fig:mod_model}
plots the velocity versus time of the top of the slinky and shows
that the top is initially at rest, then accelerates rapidly
until time $t_{\mathrm{lin}}=0.03$\,s during the initial linear 
collapse, which is marked by a sudden change in curvature of the 
velocity profile. The initial dynamics of the top differ from the
instant-collapse model; in particular the velocity of the
top of the slinky at time $t=0$ is zero, rather than having a 
finite value. 
However, after the initial acceleration of the top, the velocity 
variation of the top is similar to that in the instant-collapse model.

The right panel in Fig.~\ref{fig:slinky_hang_fall} also illustrates 
the solution of the finite-collapse-time model with the typical 
parameters, showing a helix drawn to
match $x(\xi,\frac{1}{2}t_c )$---the model slinky at one half 
the total collapse time. The upper, dark-gray
(blue online) section of the helix 
is the portion of the slinky above the collapse front,
described by Eq.~(\ref{eqn:collapsing2}).
The location of the collapse front is
shown by a dashed horizontal line, while
the dot (red online) shows the center of mass and 
the light-gray (green online) section at the bottom is the collapsed 
section in the hanging configuration.


\section{Modeling real slinkies \label{sec:appl}}

\subsection{Data \label{sec:appl_data}}

\begin{figure}[h!]
\begin{center}
\scalebox{0.45}{\includegraphics{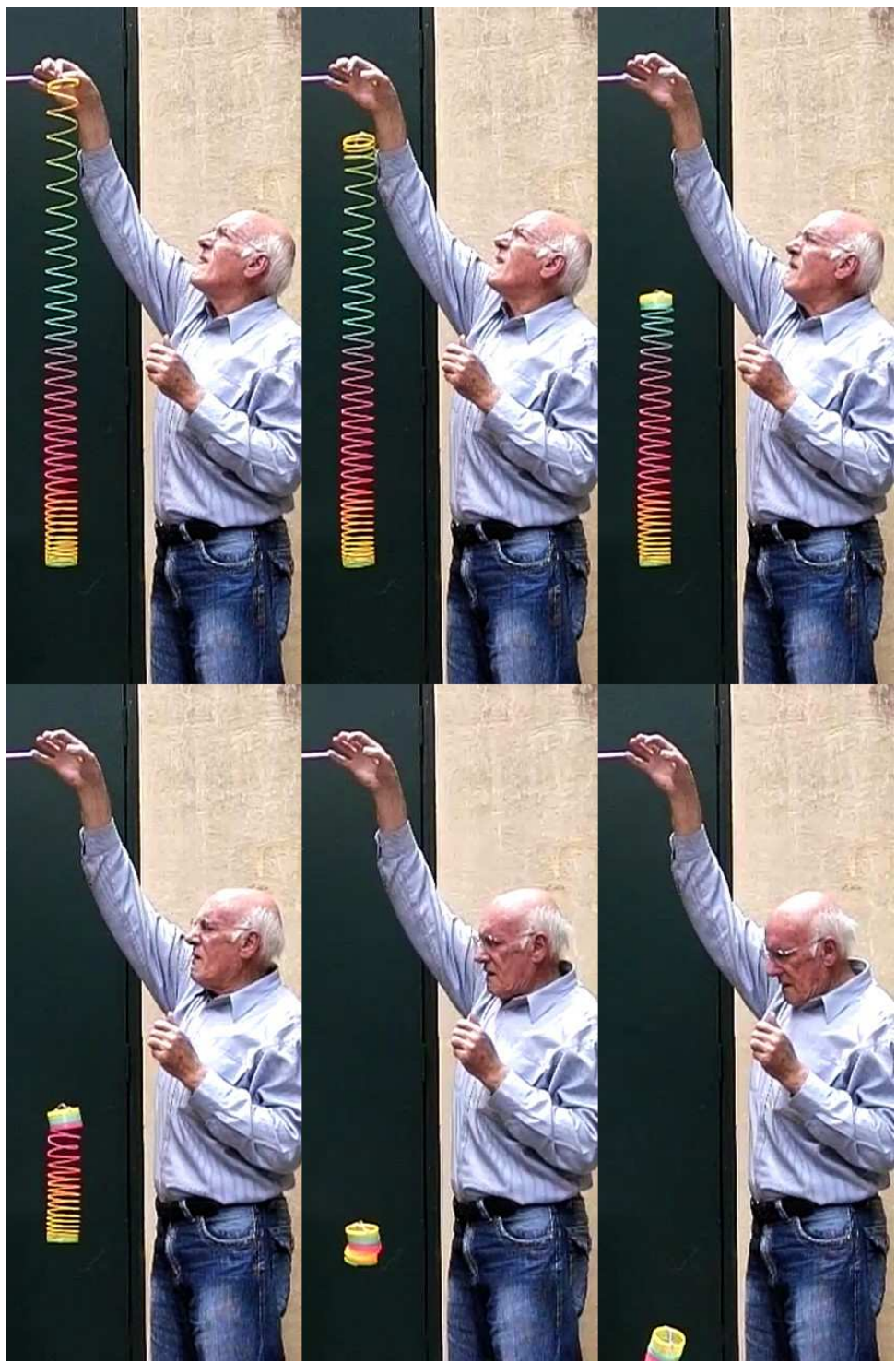}}
\caption{\label{fig:slinky_video}Frames extracted from a 
high-speed video of the fall of a rainbow-colored slinky, illustrating
the collapse of the top of the slinky, and the continued suspension
of the bottom after release of the top. The top
end of the slinky takes $\sim 0.25$\,s to reach the bottom.}
\end{center}
\end{figure}

\begin{figure}[h!]
\begin{center}
\scalebox{0.7}{\includegraphics{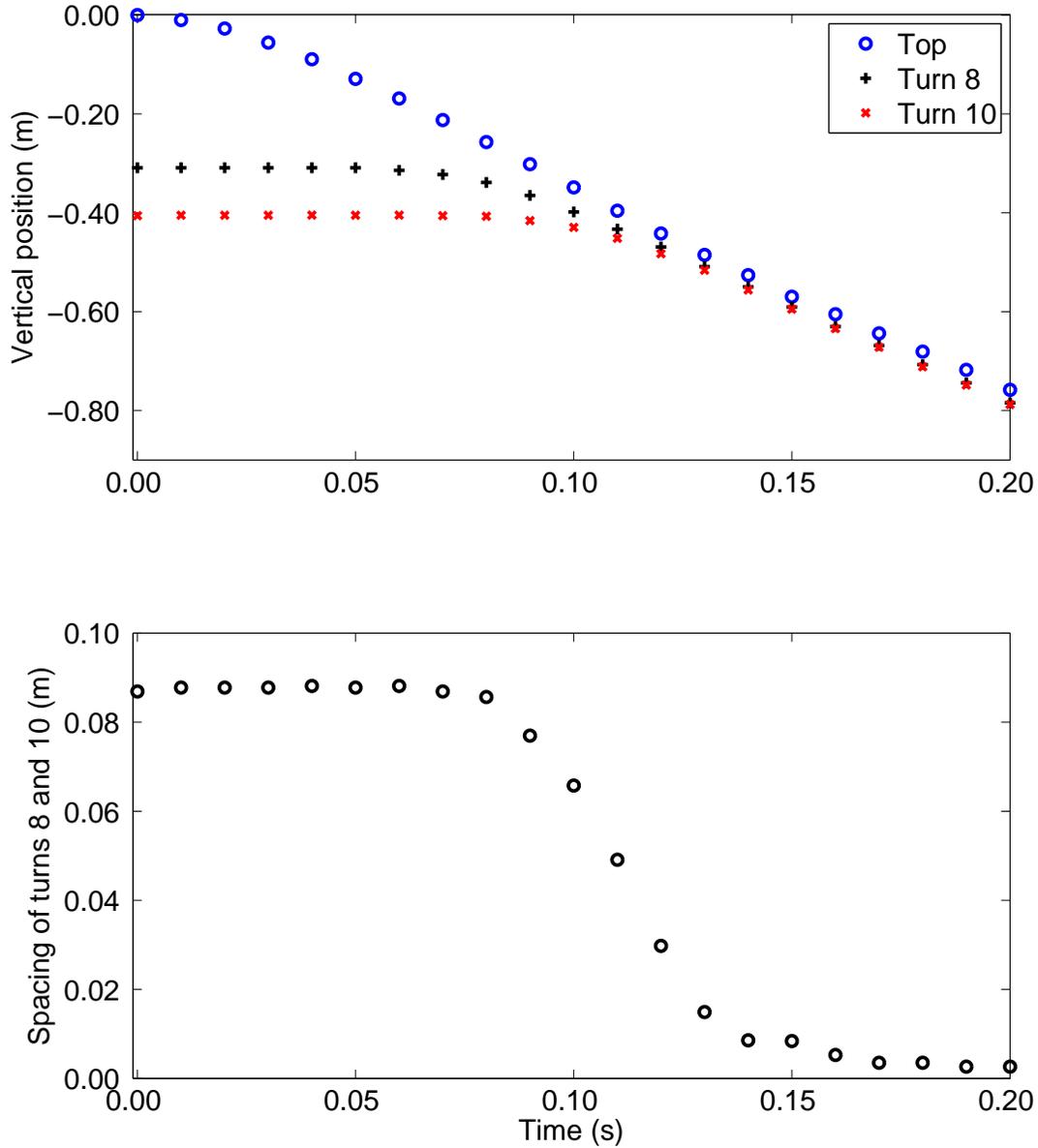}}
\caption{\label{fig:collapse_diagram}Data extracted from the 
video shown in Fig.~\ref{fig:slinky_video}, illustrating the finite
time for collapse of the turns of the slinky. Upper panel: position
versus time of the top of the slinky (circles, blue online), turn eight of the
slinky ($+$ symbols, black online) and turn ten of the slinky ($\times$
symbols, red online). Position is negative downwards in this panel. Lower panel:
The spacing of turns eight and ten versus time.}
\end{center}
\end{figure}

The finite-collapse-time model from Section~\ref{sec:model_mod} is 
compared with data obtained for two real slinkies, labeled A and B. 
The masses, 
lengths, and numbers of turns of the slinkies are 
listed in Table I. Slinky A is a typical metal slinky and slinky B 
is the light plastic rainbow-colored slinky shown in 
Fig.~\ref{fig:slinky_video}. These two slinkies were chosen because
they have significantly different parameters.

\begin{table}[!ht]
\centering
\caption{Measured data for two real slinkies.}
\label{tab:obs_data_slinkies}
\begin{tabular}{lll}
\hline\hline
& Slinky A & Slinky B\\
\hline
Mass $m$\,(g) & 215.5 & 48.7\\
Collapsed length $l_1$\,(mm) & 58 & 66 \\
Stretched length $X_{\mathrm{B}}$\,(m) & 1.26 & 1.14 \\
Number of turns $N$& 86 & 39 \\
\hline\hline
\end{tabular}
\end{table}

The slinkies are suspended from a tripod and released, and the fall
is captured with a Casio EX-F1 camera at 300 frames/s. The positions of 
the top and bottom of each slinky are determined from the movies 
at time steps of $\tau=0.01$\,s in each case. Figure~\ref{fig:slinky_video} 
shows frames from the movie used to obtain the data for slinky B. 

\subsection{Fitting the data and model \label{sec:appl_model}}

The finite-collapse time model from Section~\ref{sec:model_mod} is 
applied to the data for the two slinkies as follows. The observed 
positions for the top of each slinky during its fall are fitted to 
the model using least squares for all time steps.
The free parameters in the model are taken to be the collapse 
mass fraction $\Delta$, the spring constant $k$, and an offset $t_0$
to time, which describes the time of release of the slinky compared
to the time of the first observation. The parameter $t_0$ is 
needed because the precise time of release is difficult to 
determine accurately. The additional slinky 
parameters used are the measured values of the collapsed length 
$\ell_1$, the hanging length $X_{\mathrm{B}}$, and the mass $m$. 
(Given $\ell_1$, $X_{\mathrm{B}}$, $m$, and a chosen value of $k$, 
Eq.~(\ref{eqn:xb0}) determines the value of $\xi_1$, so equivalently, 
$\xi_1$ could be taken as a free parameter instead of $k$.) 

The method of fitting is to fit the data values 
$x_{\mathrm{T}}(t_n)$ for the positions of the top of a slinky ($T$
denotes top) at the observed times $t_n=(n-1)\tau$ (with $n=1,2,...$) 
to the model function for the positions evaluated at the 
offset time, i.e.\ the fit is made to $x(\xi,t)$ 
evaluated at $\xi=0$ and $t=t_n-t_0$. The model function 
$x(\xi,t)$ is defined by Eqs.~(\ref{eqn:collapsed3}), 
(\ref{eqn:collapsed4}), (\ref{eqn:cubic_mod1}), 
and~(\ref{eqn:cubic_mod2}) (and by the hanging configuration $X(\xi)$ 
for $t< t_0$). This procedure correctly identifies $t=t_0$ as the time 
of release.

Table~II lists the best-fit parameters for the slinkies. The value
of $\Delta$ is given both as a mass fraction and in 
terms of the corresponding number of turns of the slinky. 
For the plastic slinky $\xi_1=1$, implying 
that no turns are collapsed at the bottom of the slinky 
in the hanging configuration. Inspection of the top left frame in 
Fig.~\ref{fig:slinky_video} suggests that this is correct.

\begin{table}[!ht]
\centering
\caption{Best-fit model parameters for the slinkies.}
\label{tab:mod_params_slinkies}
\begin{tabular}{lll}
\hline\hline
& Slinky A & Slinky B\\
\hline
Spring constant $k$\,(N/m) & 0.69  & 0.22\\
$\xi_1$ & 0.89 & 1 \\
$\xi_1$\,(collapsed turns) & 9.5 & 0\\
$\Delta$ & 0.045 & 0.45 \\
$\Delta$\,(turns) & 3.9 & 18 \\
$t_0$\,(s) & 0.022 & 0.01\\
Total collapse time $t_c$\,(s) & 0.27 & 0.27 \\
Linear collapse time $t_{\mathrm{lin}}$\,(s) & 0.014 & 0.12\\
\hline\hline
\end{tabular}
\end{table}

\begin{figure}[h!]
\begin{center}
\scalebox{0.7}{\includegraphics{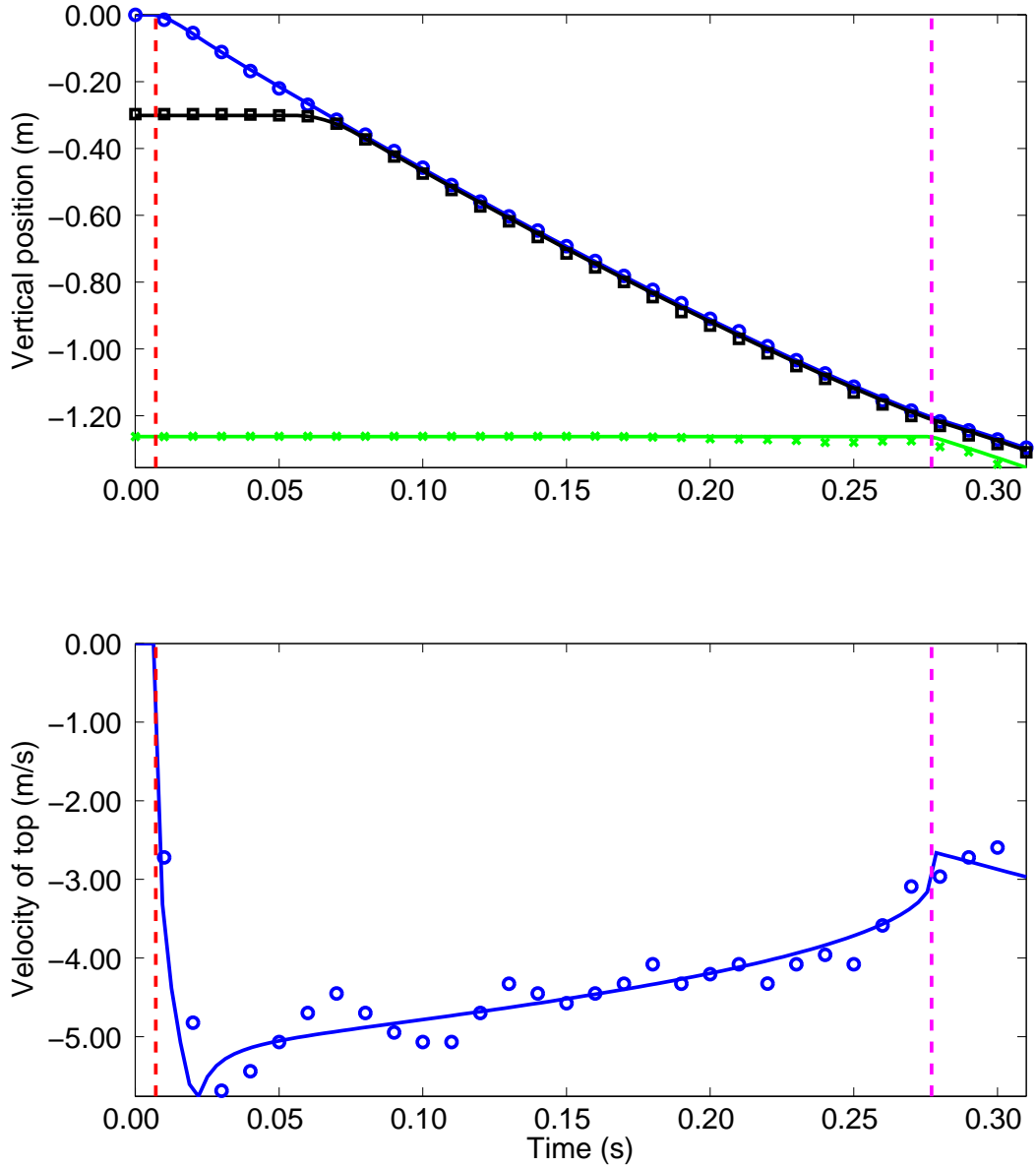}}
\caption{\label{fig:mod_4689}The finite-collapse-time model
applied to slinky A. The upper panel shows position versus time
for the slinky top (upper), turn 10 (middle),
and slinky bottom (lower),
with the observed data represented by symbols and the best-fit
model values by curves. The fitting is based on the observed positions
of the slinky top. The lower panel shows the velocity of the slinky 
top versus time. The vertical
dashed lines in both panels show the time of release of the slinky 
(left), which is a model parameter, and the model collapse time
(right).}
\end{center}
\end{figure}

\begin{figure}[h!]
\begin{center}
\scalebox{0.7}{\includegraphics{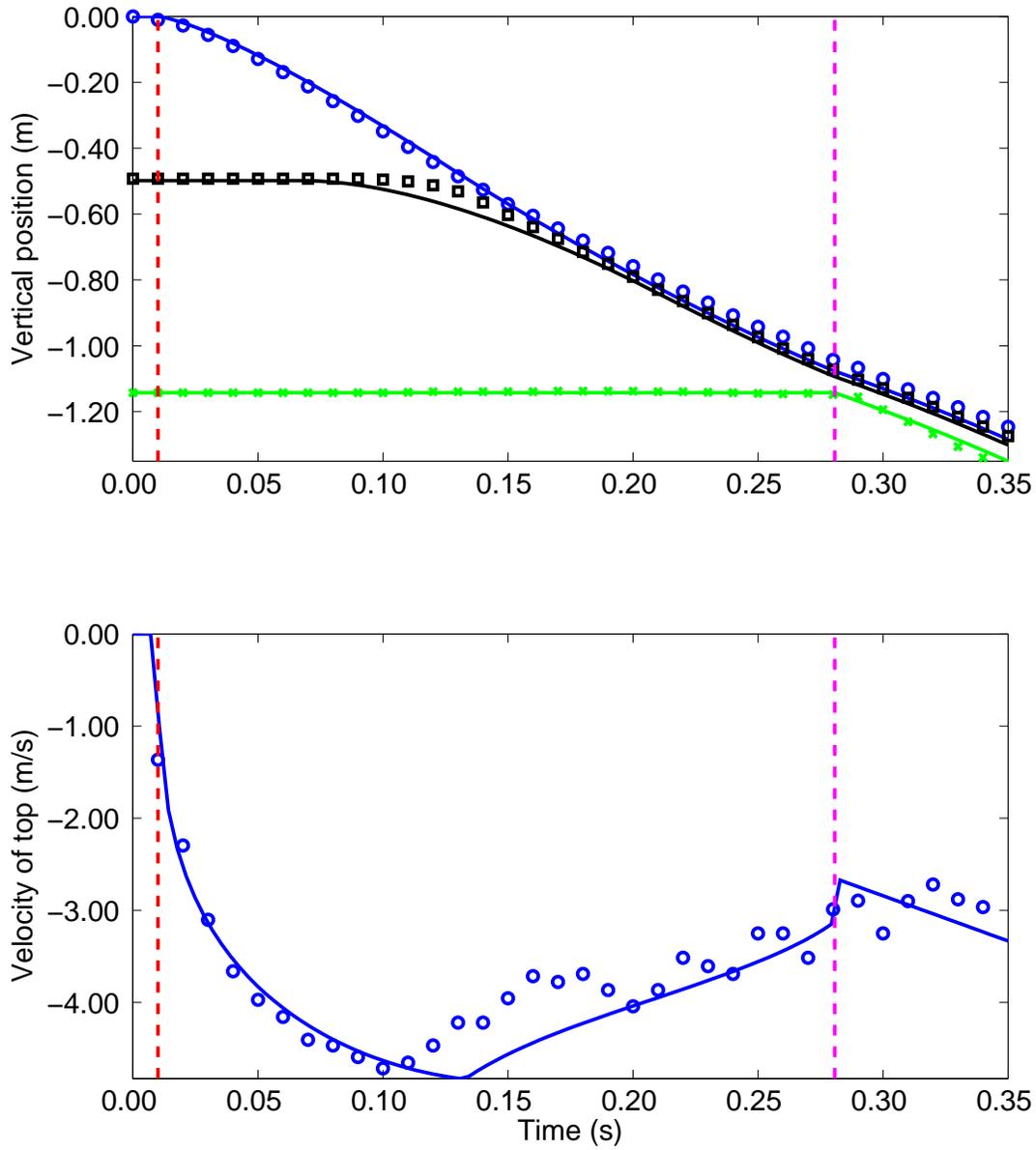}}
\caption{\label{fig:mod_4696}The finite-collapse-time model
applied to slinky B. The presentation is the same as in 
Fig.~\ref{fig:mod_4689}.}
\end{center}
\end{figure}

Figures~\ref{fig:mod_4689} and~\ref{fig:mod_4696} show the fits between 
the model and the observed data for Slinkies A and B, respectively. The upper panel in each figure shows 
positions versus time for the slinky top (model: upper solid curve, data: circles, blue online),
turn 10 (model: middle solid curve, data: squares, black online), and the slinky bottom 
(model: lower solid curve, data: $\times$, green online).
The lower panel in each figure shows the velocity of the top of 
the slinky versus time (model: solid curve, data: circles,
blue online). The measured velocity of the top of the 
slinky is determined by centered differencing of the observed 
position values, i.e.\ the velocity at time $t_n$ is approximated by
\begin{equation}\label{eqn:cdiff}
v_{\mathrm{T}}(t_n)=\frac{x_{\mathrm{T}}(t_{n+1})
  -x_{\mathrm{T}}(t_{n-1})}{2\tau}.
\end{equation}
These values are estimated for illustrative comparison with the model,
but they are not used in the fitting, which uses only the position data
for the top shown in the upper panel.
Note also that the lower panel shows downward values as negative, 
i.e.\ it shows $-v_{\mathrm{T}}(t_n)$ versus $t_n$.
Both panels in Figs.~\ref{fig:mod_4689} and~\ref{fig:mod_4696} also
show the time offset $t_0$ for the model by the left vertical (red online) dashed 
line, and the total collapse time for the model by the right vertical (pink online) 
dashed line.

The results in Figs.~\ref{fig:mod_4689} and~\ref{fig:mod_4696} 
demonstrate that the model achieves a good qualitative fit to the 
observed positions of the top of each slinky. 
The quality of the fit is shown in the approximate reproduction of 
the values of the velocity of the top of each slinky obtained by 
differencing the position data for the top. In particular, the 
description of the finite time for collapse of the slinky top given
by Eq.~(\ref{eqn:collapsing1}), with the best-fit model values,
approximately reproduces the observed initial variation in 
the velocity of the top of each slinky shown in the lower panels of 
the figures. Although we do not attempt a detailed error analysis,
it is useful to consider the expected size of uncertainties in the
data values. If the observed position values are accurate to
$\sigma_x\approx 0.5$\,cm, the uncertainty in velocity implied by
the centered differencing formula Eq.~(\ref{eqn:cdiff}) is
\begin{equation}\label{eqn:unc_v}
\sigma_v=\frac{\sigma_x}{\sqrt{2}\tau}\approx 0.4\,\mathrm{m/s}. 
\end{equation}
The detailed 
differences between the observed and best-fit model velocity values
are approximately consistent with Eq.~(\ref{eqn:unc_v}).

The fit is better for slinky A than slinky B, as shown by specific 
discrepancies between the model and observed data for the position 
of turn 10 (upper panel of Fig.~\ref{fig:mod_4696}), and the velocity 
of the top (lower panel of Fig.~\ref{fig:mod_4696}). This may be due 
to the technique used to hang the slinky: the top turns are 
tied together to allow the slinky to be hung vertically (see the first 
frame in Fig.~\ref{fig:slinky_video}). About a turn and a half of the 
slinky was joined at the top, and as a result the top of the slinky is 
heavier than in the model, and there is approximately one fewer turn. 
The same technique was used for both slinkies, but the effect may be 
more important for slinky B, which is significantly lighter and has fewer turns, 
than slinky A. We make no attempt to incorporate this in our model.

The best-fit values for the model parameter $\xi_1$ may be checked 
by comparison with the observed number of collapsed 
turns $N_c$ at the bottom of each slinky in the hanging configuration, 
which is given by Eq.~(\ref{eqn:nc}). 
Alternatively, the model values for the spring constant $k$
may be checked by comparison with the period of the fundamental mode
of oscillation of the slinky when it is hanging~\cite{young93}
\begin{equation}\label{eqn:fundamental}
T_0=4\sqrt{\frac{m}{k}}.
\end{equation}
Table~\ref{tab:pred_xi_k_slinkies} 
lists the predictions for $N_c$ and $T_0$ based on the model values 
of $\xi_1$ and $k$, and the observed values for each slinky.

\begin{table}[!ht]
\centering
\caption{Predictions (for best-fit model parameters) and observations 
for the number of collapsed turns when hanging, and for the
fundamental mode frequency.}
\label{tab:pred_xi_k_slinkies}
\begin{tabular}{lll}
\hline\hline
& Slinky A & Slinky B\\
\hline
Model fundamental period $T_0$\,(s) & 2.23 & 1.88\\
Observed fundamental period\,(s) &  2.18 & 1.77 \\
Model number of collapsed turns $N_c$ & 9.5 & 0\\
Observed number of collapsed turns & 10 & 0 \\
\hline\hline
\end{tabular}
\end{table}

Table~\ref{tab:pred_xi_k_slinkies} shows that the slinky model with
best-fit parameters approximately reproduces the observed fundamental 
mode periods and numbers of collapsed turns for the two slinkies. 
(Note that the two model values $N_c$ and $T_0$ are not independent.) 
The discrepancies between the model and observed values for the 
fundamental periods are $\sim 5$\%, with the model values being too
large in both cases. It is useful to consider the expected
size of discrepancies in the period produced by observational 
uncertainties. From Eqs.~(\ref{eqn:xb0}), 
(\ref{eqn:tc}), and~(\ref{eqn:fundamental}) it follows that
\begin{equation}\label{eqn:tc2}
t_c=\frac{4}{3}\left[\frac{2(X_{\mathrm{B}}-\ell_1)}{g}\right]^{3/4}
  \frac{1}{\sqrt{T_0}}.
\end{equation}
Assuming the distances $X_{\mathrm{B}}$ and $\ell_1$ are well-determined,
Eq.~(\ref{eqn:tc2}) implies
\begin{equation}\label{eqn:sigma_fundamental}
\frac{\sigma_{T_0}}{T_0}=2\frac{\sigma_{t_c}}{t_c},
\end{equation}
where $\sigma_{T_0}$ and $\sigma_{t_c}$ are the uncertainties in $T_0$
and $t_c$ respectively. Taking the value of the time 
step $\tau=0.01$\,s as a representative value for $\sigma_{t_c}$ in
Eq.~(\ref{eqn:sigma_fundamental}) gives 
$\sigma_{T_0}/T_{0}= 0.08$, i.e.\ an 8\% error in the model value for
the fundamental mode period. This suggests 
that the model values for $T_0$ are as accurate as might be 
expected from observational uncertainties, and indicates that it 
is difficult to determine the mode period for a real slinky 
based on measuring the fall of the slinky.

\begin{figure}[h!]
\begin{center}
\scalebox{0.35}{\includegraphics{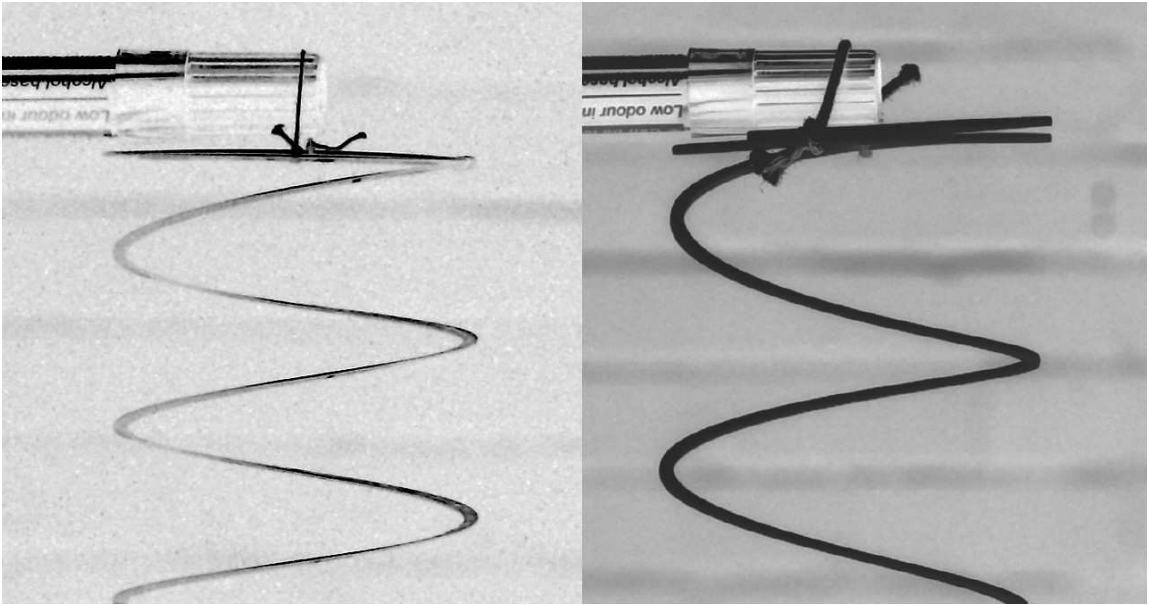}}
\caption{\label{fig:top_expt_susp}An experiment with different methods 
of suspension of the top of slinky B. In the left-hand image the top is
suspended from a string tied across a diameter of the first turn 
of the slinky. In the right-hand image the string is tied around the 
first two turns.}
\end{center}
\end{figure}

The technique of suspension of the top of the slinky, involving tying
about a turn and a half of the slinky together to ensure that it hangs
vertically, introduces some uncertainty into the modeling. It is
interesting to investigate the effect of this step on the 
initial dynamics of the slinky during the fall. For this purpose the 
slinky is suspended in two additional ways, with a string tied 
across both sides of just the top turn, and with a string tied across 
both sides of the top two turns, linked together. These methods
of suspension involve fewer, and greater numbers of turns tied together 
at the top, respectively, compared with the original method (which had
about a turn and a half tied together at the top). 
Figure~\ref{fig:top_expt_susp} illustrates the two methods of 
suspension, showing images (in inverted grayscale for clarity) of
the top few turns of the slinky in the two cases. The left-hand image
shows the case with one turn tied together at the top while the 
right-hand image shows the case with two turns tied together.

\begin{figure}[h!]
\begin{center}
\scalebox{0.7}{\includegraphics{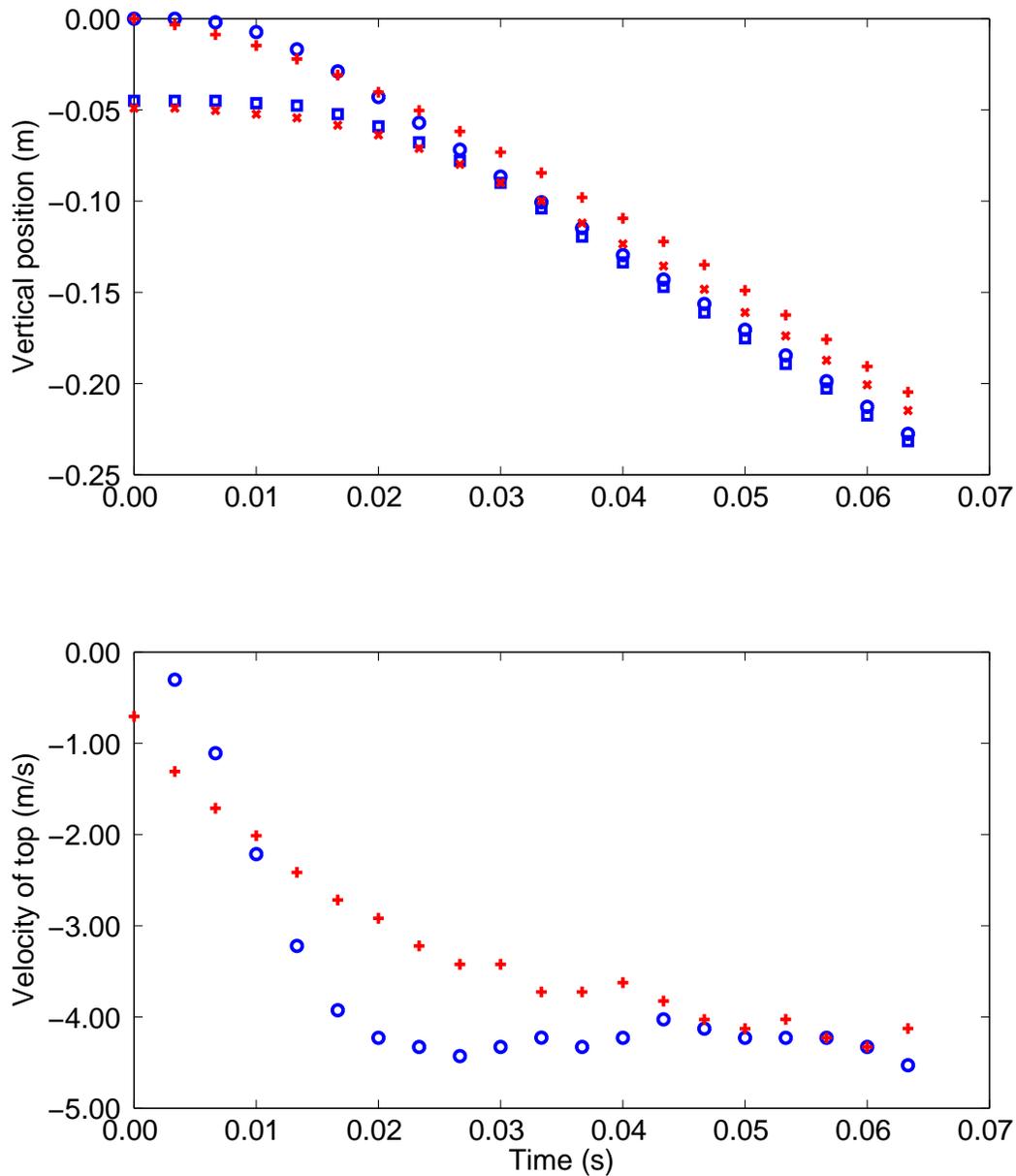}}
\caption{\label{fig:top_expt_data}Data extracted for the initial
fall of slinky B following suspension using the two methods shown in 
Fig.~\ref{fig:top_expt_susp}. The circles and squares show results for suspension
by one turn and the $+$ and $\times$ symbols for suspension by two turns. The upper panel shows
the position versus time of the top and of the first turn below the
initially tied top section. The lower panel shows the velocities 
of the top in each case, obtained by differencing
the position data (circles for suspension by one turn and $+$
symbols for suspension by two).}
\end{center}
\end{figure}

With these methods
of suspension the slinky is filmed being dropped, and data are 
extracted for the first 0.06\,s of the fall in each case; the results are
shown in Fig.~\ref{fig:top_expt_data}. The upper panel shows the 
positions versus time for the top of the slinky and for the first turn
below the turns tied at the top, for each case: circles and 
squares, respectively for suspension by one turn,
and $+$ and $\times$ symbols, respectively
for suspension by two. The lower panel shows the velocities of
the top in each case, obtained by differencing the position data using
Eq.~(\ref{eqn:cdiff}) (circles for suspension by one turn and 
$+$ symbols for suspension by two). These results show that the 
top of the slinky accelerates from rest more rapidly when fewer turns
are tied together at the top, which is expected because the inertia
of the top is reduced. However, in both cases the top achieves a very
similar velocity after $\sim 0.05$\,s. It is expected that
the subsequent dynamics of  the collapse of turns will be similar in
the two cases. The dependence of the initial dynamics of the top on
the method of suspension will influence the estimates of model
parameters, in particular the collapse mass fraction $\Delta$.
However, it is expected that the estimate of the spring constant $k$
will be less influenced because this parameter is determined largely 
by the identification from the data of the total collapse time $t_c$. 
The dependence of the fitting on the  method of suspension of the top 
of the slinky could perhaps be reduced by fitting to the positions of
turns other than the top turn during the fall.


\section{Conclusions \label{sec:conc}}

The fall of a slinky illustrates the physics of a tension spring,
and more generally wave propagation in a spring. This paper 
investigates the dynamics of an initially stretched slinky that is 
dropped. During the fall the slinky turns collapse from the top
down as a wave front propagates along the slinky. The bottom of 
the slinky does not begin to fall until the top collides with it. 
A modification to an existing model~\cite{calkin93} for the fall is 
presented, providing an improved description of the collapse of the
slinky turns. The modification is the inclusion of a finite 
time for collapse of turns behind the downward propagating wave
front. The new model is fitted to data obtained from videos of the
falls of two real slinkies having different properties.

The model is shown to account for the observed positions of the top of
each slinky in the experiments, and in particular reproduces the 
initial time-profile
for the velocity of the top after release. The spring constant of the 
slinky is assumed as a free parameter in the model, and the best-fit 
model values are tested by comparison with independent determinations 
of the fundamental mode periods for the two slinkies, which depend on 
the spring constants. The model values appear consistent with the 
observations taking into account the observational uncertainties.

The new model for the slinky dynamics during the fall developed here is  
semi-analytic, and allows treatment of a tension spring including
approximate description of the dynamics of the collapse of the
spring. During the collapse of the top of the slinky the turns collide,
but the model does not describe this process in detail. Instead,
the collapse is approximately described by the assumption of a linear
decrease in tension as a function of mass density along the spring 
behind the front initiating the collapse. The linear approximation 
is motivated by the experimental data from the slinky videos, which 
shows that the spacing between slinky turns during the collapse 
decreases approximately linearly with time. 

The behavior of a falling slinky is likely to be counter-intuitive to 
students and provides a useful (and very simple) undergraduate
physics lecture demonstration. The explanation of the behavior may be
supplemented by showing high-speed videos of the fall. The 
modeling of the process presented here is also relatively simple, and 
should be accessible to undergraduate students.




\newpage

\end{document}